\documentclass[pdftex,cite,aps,prl,superscriptaddress,amsmath,amsfonts,amssymb,twocolumn,showpacs]{revtex4}

\renewcommand{\vec}[1]{\mathbf{#1}} 

\usepackage{amsmath}
\usepackage{amsfonts}
\usepackage{amssymb}
\usepackage[pdftex]{graphicx}

\newcommand{\citeasnoun}[1]{Ref.~\onlinecite{#1}}

\newcommand{\tauk}[2]{\tau_{#1,#2}}
\newcommand{\E}[2]{E_{#1 #2}}
\newcommand{\chiany}[1]{\chi^{(#1)}}
\newcommand{\chitwo}{\chiany{2}}
\newcommand{\chithree}{\chiany{3}}
\newcommand{\chiell}{\chiany{\ell}}
\renewcommand{\eqref}[1]{Eq.~\ref{#1}}

\renewcommand{\Re}{\operatorname{Re}}
\renewcommand{\Im}{\operatorname{Im}}

\begin{document}
\def\linefigwidth{0.5\textwidth}
\def\smalllinefigwidth{0.35\textwidth}
\def\smallerlinefigwidth{0.25\textwidth}
\def\largelinefigwidth{0.5\textwidth}
\def\figref{Fig.~\ref}

\title{$\chitwo$ and $\chithree$ harmonic generation at a critical power \\ in inhomogeneous doubly resonant cavities}

\author{Alejandro Rodriguez, Marin Solja{\v{c}}i{\'{c}}, J. D.
Joannopoulos, and Steven~G.~Johnson} 

\address{Center for Materials
Science and Engineering, Massachusetts Institute of Technology,
Cambridge, MA 02139}

\email{alexrod7@mit.edu}

\begin{abstract}
  We derive general conditions for 100\% frequency conversion in any
  doubly resonant nonlinear cavity, for both second- and
  third-harmonic generation via $\chitwo$ and $\chithree$
  nonlinearities.  We find that conversion efficiency is optimized for
  a certain ``critical'' power depending on the cavity parameters, and
  assuming reasonable parameters we predict 100\% conversion using
  milliwatts of power or less.  These results follow from a
  semi-analytical coupled-mode theory framework which is generalized
  from previous work to include both $\chitwo$ and $\chithree$ media
  as well as inhomogeneous (fully vectorial) cavities, analyzed in
  the high-efficiency limit where down-conversion processes lead to a
  maximum efficiency at the critical power, and which is verified by
  direct finite-difference time-domain (FDTD) simulations of the
  nonlinear Maxwell equations.  Explicit formulas for the nonlinear
  coupling coefficients are derived in terms of the linear cavity
  eigenmodes, which can be used to design and evaluate cavities in
  arbitrary geometries.
\end{abstract}

\maketitle

%\ocis{(190.2620) Nonlinear optics: frequency conversion; (230.4320) Nonlinear optical devices}

%%%%%%%%%%%%%%%%%%%%%%%%%% REFS %%%%%%%%%%%%%%%%%%%%%%%%%%%%%%%
%\bibliographystyle{osajnl}
%\bibliography{photon}
%%%%%%%%%%%%%%%%%%%%%%%%% body %%%%%%%%%%%%%%%%%%%%%%%%%%%%%%%

\section{Introduction}

In this paper, we consider second- and third-harmonic generation in
doubly resonant cavities.  We generalize previous experimental and
theoretical work on this subject, which had focused only on $\chitwo$
nonlinearities in large Fabry-Perot etalons or two-dimensional ring resonators
where a scalar approximation
applied~\cite{Drummond80,Wu87,Ou93,Paschotta94,Berger96,Zolotoverkh00,Maes05,Liscidini06, Dumeige06},
to incorporate both $\chitwo$ and $\chithree$ nonlinearities and
handle the inhomogenous fully vectorial case.  We then develop several
results from this generalization: whereas it is well known that 100\%
harmonic conversion is possible, at least in $\chitwo$ media, in this
work we further explore and identify the conditions under which this
can be achieved.  First, we demonstrate the existence of a critical
input power at which harmonic generation is maximized, in contrast to
previous work that focused largely on the low-power limit in which
generation efficiency increased monotonically with input
power~\cite{Ou93}.  Second, while it is well known that harmonic
conversion can be achieved at arbitarily low powers given sufficiently
long cavity lifetimes, this implies a narrow bandwidth---we show that,
by combining moderate lifetimes (0.1\% bandwidth) with tight spatial
confinement, 100\% second- and third-harmonic conversion can
theoretically be achieved with sub-milliwatt power levels.  Such
low-power conversion could find applications such as high-frequency
sources~\cite{Berger96}, ultracompact-coherent optical
sources~\cite{Moore95, Liscidini04, Murray06},
imaging~\cite{Scotto03}, and spectroscopy~\cite{McConnell01}.

Nonlinear frequency conversion has been commonly realized in the
context of waveguides~\cite{Dutta98, Aguanno02, Cowan02, Malvezzi03},
or even for free propagation in the nonlinear materials, in which
light at one frequency co-propagates with the generated light at the
harmonic frequency~\cite{Pearl99,Balakin01,Aguanno01,Norton02}.  A
phase-matching condition between the two frequencies must be satisfied
in this case in order to obtain efficient conversion~\cite{Berger96,
Dumeige06}. Moreover, as the input power is increased, the frequency
conversion eventually saturates due to competition between up and down
conversion.  Frequency conversion in a doubly resonant cavity has
three fundamental differences from this familiar case of propagating
modes.  First, light in a cavity can be much more intense for the same
input power, because of the spatial (modal volume $V$) and temporal
(lifetime $Q$) confinement.  We show that this enhances
second-harmonic ($\chitwo$) conversion by a factor of $Q^3/V$ and
enhances third-harmonic ($\chithree$) conversion by a factor of
$Q^2/V$.  Second, there is no phase-matching condition per se for
100\% conversion; the only absolute requirement is that the cavity
support two modes of the requisite frequencies.  However, there is a
constant factor in the power that is determined by an overlap integral
between the mode field patterns; in the limit of a very large cavity,
this overlap integral recovers the phase-matching condition for
$\chitwo$ processes.  Third, the frequency conversion no longer
saturates---instead, it peaks (at 100\%, with proper design) for a
certain critical input power satisfying a resonant condition, and goes
to \emph{zero} if the power is \emph{either} too small or too large.

Second-harmonic generation in cavities with a single resonant mode at
the pump
frequency~\cite{Armstrong62,Ashkin66,Smith70,Ferguson77,Brieger81,Berquist82,Kozlovsky88,Dixon89,Collet90,
Persaud90,Moore95,Schneider96,Mu01,Hald01,McConnell01,Dolgova02,Liu05,Scaccabarozzi06}
or the harmonic frequency~\cite{DiFalco06} requires much higher power
than a doubly resonant cavity, approaching one
Watt~\cite{Ou93,Scaccabarozzi06} and/or requiring amplification within
the cavity. (A closely related case is that of sum-frequency
generation in a cavity resonant at the two frequencies being
summed~\cite{Schnitzler02}.)  Second-harmonic generation in a doubly
resonant cavity, with a resonance at both the pump and harmonic
frequencies, has most commonly been analyzed in the low-efficiency
limit where nonlinear down-conversion can be
neglected~\cite{Paschotta94,Berger96,Zolotoverkh00,Maes05,Liscidini06,Dumeige06},
but down-conversion has also been included by some
authors~\cite{Drummond80, Wu87, Ou93}.  Here, we show that
not only is down-conversion impossible to neglect at high conversion
efficiencies (and is, in fact, necessary to conserve energy), but also
that it leads to a critical power where harmonic conversion is
maximized.  This critical power was demonstrated numerically by
\citeasnoun{Ren04} in a sub-optimal geometry where 100\% efficiency is
impossible, but does not seem to have been clearly explained
theoretically; the phenomenon (for $\chitwo$) was also implicit in the
equations of \citeasnoun{Ou93} but was not identified, probably
because it occurred just beyond the range of power considered in that
work.
\begin{figure}[hbt]
\centering\includegraphics[width=7cm]{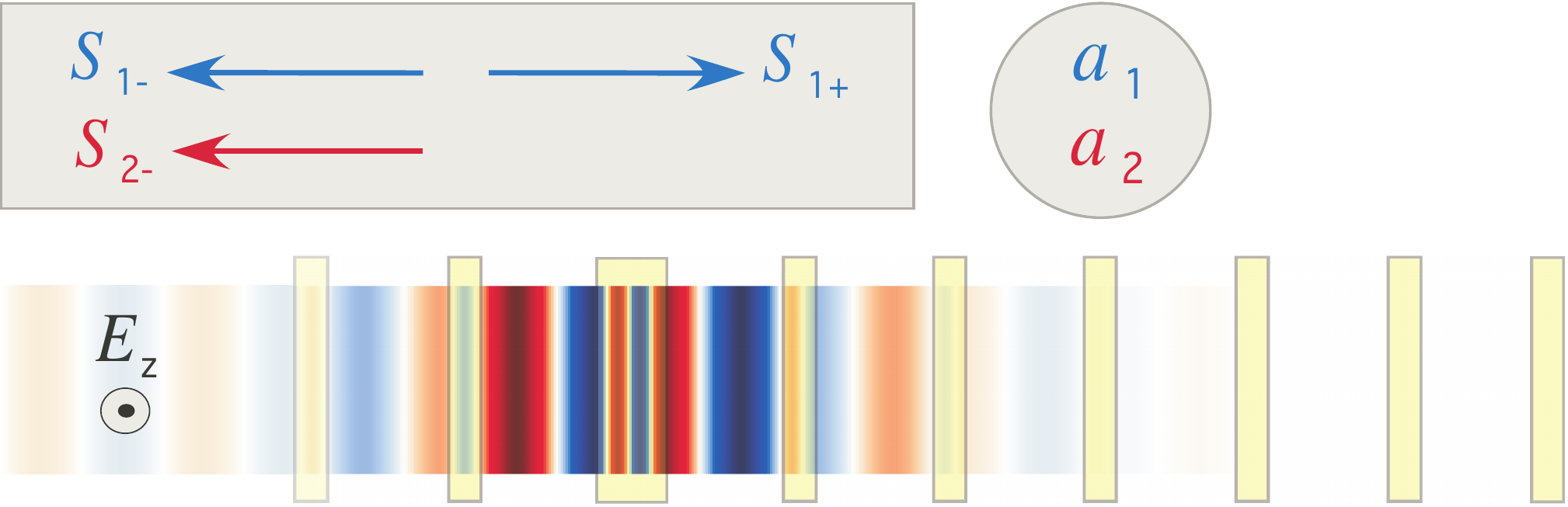}
\caption{\textit{Top:} Schematic diagram of waveguide-cavity
system. Input light from a waveguide (left) at one frequency
(amplitude $s_{1+}$) is coupled to a cavity mode (amplitude $a_1$),
converted to a cavity mode at another frequency (amplitude $a_2$) by a
nonlinear process, and radiated back into the waveguide (amplitude
$s_{2-}$).  Reflections at the first frequency ($s_{1-}$) may also
occur.  \textit{Bottom:} 1d example, formed by quarter-wave defect in
a quarter-wave dielectric stack.  Dielectric material is yellow, and
electric field $E_z$ of third-harmonic mode is shown as blue/white/red
for positive/zero/negative amplitude.}
\label{fig1}
\end{figure}

Previous work on third-harmonic generation in cavities considered only
singly resonant cavities; moreover, past work focused on the case of
$\chitwo$ materials where $3\omega$ is generated by cascading two
nonlinear processes (harmonic generation and frequency summing)
\cite{Koch99,McConnell01}. Here, we examine third-harmonic generation
using $\chithree$ materials so that only a single resonant process
need be designed and a different set of materials becomes available.
($\chithree$ third-harmonic generation in a bulk periodic structure,
with no cavity, was considered in \citeasnoun{Markowicz04}.)  In a
$\chithree$ medium, there are also self/cross-phase modulation
phenomena (nonlinear frequency shifts) that, unchecked, will prevent
100\% conversion by making the frequency ratio $\neq 3$. To address
this mismatch, we describe how one can use two materials with
opposite-sign $\chithree$ to cancel the frequency-shifting effect; it
may also be possible to pre-shift the cavity resonant frequency to
correct for the nonlinear shift.  On the other hand, a $\chitwo$
medium has no self-phase modulation, and so in this case it is
sufficient to increase the input power until 100\% frequency
conversion is reached.  (An ``effective'' self-phase modulation occurs
in $\chitwo$ media due to cascaded up- and down-conversion
processes~\cite{Stegeman93}, but these processes are fully taken into
account by our model.  We also consider media with simultaneous
$\chitwo$ and $\chithree$ nonlinearities, and show that the latter can
be made negligible.) If the critical field were too intense, then
material breakdown might also be an obstacle, but we show that it is
sufficient to use modes with a large lifetime $Q$ and small volume $V$
so that a slow conversion due to a weak nonlinear effect has enough
time to occur.

In particular, we consider the general situation depicted
schematically in Fig.~\ref{fig1}: a two-mode nonlinear cavity coupled
to an input/output channel.  For example, a one-dimensional
realization of this is shown in Fig.~\ref{fig1}: a Fabry-Perot cavity
between two quarter-wave stacks~\cite{Joannopoulos95}, where the stack
has fewer layers on one side so that light can enter/escape.  For a
nonlinear effect, we consider specifically a $\chiell$ nonlinearity,
corresponding essentially to a shift in the refractive index
proportional to the nonlinear susceptibility $\chiell$ multiplied by
electric field $\vec{E}$ to the $(\ell-1)^{th}$ power.  Most commonly,
one would have either a $\chitwo$ (Pockels) or $\chithree$ (Kerr)
effect.  Such a nonlinearity results in harmonic generation
~\cite{Boyd92}: light with frequency $\omega$ is coupled to light with
frequency $\ell\omega$.  Therefore, if we design the cavity so that it
supports two modes, one at $\omega$ and one at $\ell\omega$, then
input power at $\omega$ can be converted, at least partially, to
output power at $\ell\omega$.

In the following, we derive a semi-analytical description of harmonic
generation using the framework of coupled-mode
theory~\cite{Drummond80,Wu87,Yariv88,Collet90,Moore95,Berger96,McConnell01,Ou93,Liu05, Dumeige06},
and then check it via direct numerical simulation of the nonlinear
Maxwell equations~\cite{Bethune89, Hashizume95, Maes05}.  For maximum
generality, we derive the coupled-mode equations using two
complementary approaches.  First, we use ``temporal'' coupled-mode
theory~\cite{Haus84:coupled, Suh04}, in which the general form of the
equations is determined only from principles such as conservation of
energy and reciprocity, independent of the specific physical problem
(for example, electromagnetic or acoustic waves).  Second, we apply
perturbation theory directly to Maxwell's equations in order to obtain
the same equations but with specific formulas for the coupling
coefficients in terms of the linear eigenmodes.  Unlike most previous
treatments of this problem~\cite{Yariv88,Drummond80,Wu87}, we do not
make a one-dimensional or scalar approximation for the electromagnetic
fields (invalid for wavelength-scale cavities), and we consider both
$\chitwo$ and $\chithree$ media.  (The optimization of these coupling
coefficients is then the generalization of the phase-matching criteria
used in one-dimensional geometries~\cite{Berger96}.)

\section{Temporal coupled-mode theory}

We derive coupled-mode equations describing the interaction of light
in a multi-mode cavity filled with nonlinear material and coupled to
input/output ports, from which light can couple in ($s_{+}$) and out
($s_{-}$) of the cavity. A schematic illustration of the system is
shown in Fig.~\ref{fig1}.  Specifically, we follow the formalism
described in \citeasnoun{Haus84:coupled}, adapted to handle
nonlinearly coupled modes with frequencies $\omega_k$.  Although
similar equations for the case of $\chitwo$ media were derived in the
past~\cite{Ou93}, they do not seem to have been derived for
$\chithree$ harmonic generation in cavities. Moreover, a derivation
via the temporal coupled-mode formalism of \citeasnoun{Haus84:coupled}
is arguably more general than earlier developments based on a
particular scalar nonlinear wave equation, because this formalism (for
a given-order nonlinearity) depends only on general considerations
such as weak coupling and energy conservation (the resulting equations
hold for vector or scalar waves in electromagnetism, acoustics, or any
other weakly-coupled problem with a few simple properties).  In the
next section, we will then specialize the equations to
electromagnetism by deriving explicit equations for the coupling
coefficients from Maxwell's equations.

We let $a_k$ denote the time-dependent complex amplitude of the $k$th
mode, normalized so that $|a_k|^2$ is the electromagnetic energy
stored in this mode.  We let $s_{\pm}$ denote the time-dependent
amplitude of the incoming ($+$) or outgoing ($-$) wave, normalized so
that $|s_{\pm}|^2$ is the power. (More precisely, $s_\pm(t)$ is
normalized so that its Fourier transform $|\tilde{s}_\pm(\omega)|^2$
is the power at $\omega$.  Later, we will let $s_{k\pm}$ denote the
input/output power at $\omega_k$.) [In 1d, the units of
$|a_k|^2$ and $|s_{\pm}|^2$ are those of energy and power per unit
area, respectively. More generally, in $d$ dimensions, the units of
$|a_k|^2$ and $|s_{\pm}|^2$ are those of energy and power per
$\textrm{length}^{3-d}$.]  By itself, a linear cavity mode decaying
with a lifetime $\tau_k$ would be described by $da_k/dt = (i\omega_k -
1/\tau_k)a_k$. [Technically, such a decaying mode is not a
true eigenmode, but is rather a ``leaky mode''~\cite{Snyder83},
corresponding to a ``quasi-bound state'' in the Breit-Wigner
scattering theory~\cite{Landau:QM}.]  The decay rate $1/\tau_k$ can be
decomposed into $1/\tau_k = 1/\tauk{e}{k} + 1/\tauk{s}{k}$ where
$1/\tauk{e}{k}$ is the ``external'' loss rate (absorption etc.) and
$1/\tauk{s}{k}$ is the decay rate into $s_-$. When the weak coupling
($\omega_k \tau_k \gg 1$) to $s_\pm$ is included, energy conservation
and similar fundamental constraints lead to equations of the
form~\cite{Haus84}:

\begin{align}
\frac{da_k}{dt} &= \left( i\omega_k - \frac{1}{\tau_k} \right) a_k
                     + \sqrt\frac{2}{\tauk{s}{k}} s_+
\label{eq1}
\\
s_- &= -s_+ + \sqrt\frac{2}{\tauk{s}{k}} a_k
\end{align}

This can be generalized to incorporate multiple input/output ports,
direct coupling between the ports, and so on~\cite{Suh04}.  The only
unknown parameters in this model are then the frequencies $\omega_k$
and the decay rates $1/\tau_k$, which can be determined by any
numerical method to solve for the cavity modes (e.g. FDTD, below).
Instead of $\tau_k$, one commonly uses the quality factor $Q_k =
\omega_k \tau_k / 2$.

Nonlinearity modifies this picture with two new amplitude-dependent
effects: a shift in the frequency (and decay rate) of the cavity, and
a coupling of one cavity mode to another.  We neglect nonlinear
effects on the input/output ports, under the assumption that intense
fields are only present in the cavity (due to spatial and temporal
confinement).  We will also make two standard assumptions of nonlinear
systems.  First, that the nonlinearities are weak, in the sense that
we can neglect terms of order $(\chiell)^2$ or higher; this is true in
practice because nonlinear index shifts are always under 1\% lest
material breakdown occur.  Second, we make the rotating wave
approximation: since the coupling is weak, we only include terms for
$a_k$ that have frequency near $\omega_k$.  In particular, we suppose
that $\omega_k \approx k \omega_1$, so that $\omega_k$ is the $k$th
harmonic.  The result is that, for a given order nonlinearity, there
are only a few possible new terms that can appear in the coupled-mode
equations.  In particular, for a $\chitwo$ nonlinearity with two modes
$\omega_1$ and its second harmonic $\omega_2$, the coupled-mode
equations must take the form:
\begin{eqnarray}
  \frac{da_1}{dt} & = & \left(i\omega_1- \frac{1}{\tau_1}\right) a_1-i\omega_1\beta_1 a^{*}_1 a_2 + \sqrt{\frac{2}{\tauk{s}{1}}}s_{+}
  \label{chi2a}
  \\ \frac{da_2}{dt} & = & \left(i\omega_2 - \frac{1}{\tau_2} \right)
  a_2 -i\omega_2\beta_2 a^2_1 + \sqrt{\frac{2}{\tauk{s}{2}}}s_{+}
  \label{chi2b}
\end{eqnarray}
Similarly, for a $\chithree$ nonlinearity with two modes
$\omega_1$ and its third harmonic $\omega_3$, the coupled-mode
equations must take the form:
\begin{widetext}
\begin{eqnarray}
  \frac{da_1}{dt} & = & \left(i\omega_1\big(1-\alpha_{11}\left|a_1\right|^2-\alpha_{13}\left|a_3\right|^2\big)-\frac{1}{\tau_1}\right)a_1 -i\omega_1\beta_1 (a^*_1)^2a_3 + \sqrt{\frac{2}{\tauk{s}{1}}}s_{+} 
  \label{chi3a}
  \\
  \frac{da_3}{dt} & = & \left(i\omega_3\left(1-\alpha_{33}\left|a_3\right|^2-\alpha_{31}\left|a_1\right|^2\right)-\frac{1}{\tau_3}\right) a_3 -i\omega_3\beta_3 a^3_1 + \sqrt{\frac{2}{\tauk{s}{3}}}s_{+}
  \label{chi3b}
\end{eqnarray}
\end{widetext}
In equations~\ref{chi3a}--\ref{chi3b}, one sees two kinds of
terms. The first are \emph{frequency-shifting} terms, with
coefficients $\alpha_{ij}$, dependent on one of the field
amplitudes. For $\chithree$, this effect is known as self-phase and
cross-phase modulation, which is absent for $\chitwo$ (under the
first-order rotating-wave approximation).  The second kind of term
\emph{transfers energy} between the modes, with coupling coefficients
$\beta_i$, corresponding to four-wave mixing for $\chithree$.
Furthermore, we can constrain the coupling terms $\beta_i$ by energy
conservation: $\frac{d}{dt} (\left|a_1\right|^2 + \left|a_2\right|^2 )
= 0$. For $\chitwo$, the constraint that follows is: $\omega_1\beta_1
= \omega_2\beta^*_2$; for $\chithree$, the constraint is
$\omega_1\beta_1 = \omega_3\beta^*_3$.  (This constraint holds even in
cavities with external loss as discussed in Sec.~\ref{sec:loss}:
energy is still conserved in the sense that the input power must equal
the output power plus the loss power, and so the harmonic conversion
term must lead to an equal energy loss and gain at $\omega_1$ and
$\omega_{2,3}$, respectively.)

The general process for construction of these coupled-mode equations
is as follows. The underlying nonlinearity must depend on the
physical, real part of the fields, corresponding to $(a_k + a^*_k)/2$.
It then follows that the $\chiell$ term will have $\ell$ powers of
this real part, giving various product terms like $a^*_1 a_2$ (for
$\chitwo$) and $a^*_1 a_1 a_1$ (for $\chithree$).  Most of these
terms, however, can be eliminated by the rotating-wave approximation.
In particular, we assume that each $a_k$ term is proportional to $e^{k
i\omega}$ multiplied by a slowly varying envelope, and we discard any
product term whose total frequency differs from $k\omega$ for the
$da_k/dt$ equation.  Thus, a term like $a^*_1 a_3 a_3$ would be
proportional to $e^{5i\omega}$, and would only appear in a $da_5/dt$
equation. (We focus on the simpler case of doubly resonant cavities in
this paper.)

At this point, the equations are already useful in order to reason
about what types of qualitative behaviors are possible in general.  In
fact, they are not even specific to electromagnetism and would also
apply to other situations such as acoustic resonators.  However, in
order to make quantitative predictions, one needs to know the
nonlinear coefficients $\alpha_{ij}$ and $\beta_i$ (as well as the
linear frequencies and decay rates).  The evaluation of these
coefficients requires a more detailed analysis of Maxwell's equations
as described below.

\section{Perturbation theory and coupling coefficients}

In this section, we derive explicit formulas for the nonlinear
coupling coefficients in the coupled-mode theory of the previous
section, applied to the case of electromagnetism.  Unlike previous
work, our expressions apply to the fully vectorial equations, valid
for high index-contrast materials, and we derive the $\chithree$ case
as well as $\chitwo$.  Our derivation is closely related to that of
\citeasnoun{Soljacic02:bistable}, which only considered the frequency
shifting (self-phase modulation) and not harmonic generation.

When a dielectric structure is perturbed by a small
$\delta\varepsilon$, a well-known result of perturbation theory states
that the corresponding change $\delta\omega$ in an eigenfrequency
$\omega$ is, to first order~\cite{Joannopoulos95}:
\begin{equation}
  \frac{\delta\omega}{\omega} = 
- \frac{1}{2}\frac{\int d^3\vec{x} \,
  \delta\varepsilon |\vec{E}|^2}{\int d^3\vec{x} \, \varepsilon
  \left|\vec{E}\right|^2}
= - \frac{1}{2}\frac{\int d^3\vec{x} \,
  \vec{E}^* \cdot{\delta \vec{P}}}{\int d^3\vec{x} \, \varepsilon
  \left|\vec{E}\right|^2}
\label{genpert}
\end{equation}
where $\vec{E}$ is the unperturbed electric field and $\delta\vec{P} =
\delta\varepsilon\vec{E}$ is the change in polarization density due to
$\delta\varepsilon$. In fact, Eq.~\ref{genpert} is general enough to
be used with any $\delta \vec{P}$, including the polarization that
arises from a nonlinear susceptibility. In particular, we can use it to
obtain the coupling coefficients of the CMT.

To do so, we first compute the nonlinear first-order frequency
perturbation due to the total field $\vec{E}$ present from all of the
modes. Once the frequency perturbations $\delta\omega_k$ are known, we
can re-introduce these into the coupled-mode theory by simply setting
$\omega_k \rightarrow \omega_k + \delta\omega_k$ in Eq.~\ref{eq1}.  By
comparison with Eqs.~\ref{chi2a}--\ref{chi3b}, we can then identify
the $\alpha$ and $\beta$ coefficients.

We consider first a $\chitwo$ nonlinearity, with the nonlinear
polarization $\delta\vec{P}$ given by $\delta P_i = \sum_{ijk}
\varepsilon \chitwo_{ijk}E_jE_k$, in a cavity with two modes
$\vec{E}_1$ and $\vec{E}_2$.  As before, we require that the modes
oscillate with frequency $\omega_1$ and $\omega_2 \approx 2 \omega_1$,
respectively. Taking $\vec{E} = \Re[\vec{E}_1 e^{i\omega_1t} +
\vec{E}_2 e^{i\omega_2t}]$ and using the rotating-wave approximation,
we can separate the contribution of $\delta\vec{P}$ to each
$\delta\omega_k$, to obtain the following frequency perturbations:
\begin{align}
  \frac{\delta\omega_1}{\omega_1} &= -\frac{1}{4} \frac{\int d^3\vec{x}
    \hspace{0.02 in} \sum_{ijk} \varepsilon \chitwo_{ijk} \hspace{0.02 in}
    \left[ \E{1}{i}^* \big(\E{2}{j}\E{1}{k}^*+\E{1}{j}^*\E{2}{k}\big)
    \hspace{0.02 in} \right]}{\int d^3\vec{x} \varepsilon
    \left|\vec{E}_1\right|^2 } \\ \frac{\delta\omega_2}{\omega_2} &= -
    \frac{1}{4}\frac{\int d^3\vec{x} \hspace{0.02 in} \sum_{ijk}
    \varepsilon \chitwo_{ijk} \hspace{0.02 in}
    \E{2}{i}^*\E{1}{j}\E{1}{k}}{\int d^3\vec{x} \hspace{0.02 in}
    \varepsilon \left|\vec{E}_2\right|^2} \\ \nonumber
\end{align}
Similarly, for a centro-symmetric $\chithree$ medium, $\delta\vec{P}$ is
given by $\delta\vec{P} = \varepsilon \chithree |\vec{E}|^2 \vec{E}$, with
$\vec{E} = \Re[\vec{E}_1 e^{i\omega_1t} + \vec{E}_3 e^{i\omega_3t}]$.
We obtain the following frequency perturbations:
\begin{widetext}
\begin{eqnarray}
    \frac{\delta\omega_1}{\omega_1} &=& -\frac{1}{8} \left[ \frac{\int d^3\vec{x} \varepsilon \chithree \hspace{0.02 in} \left( \left|\vec{E}_1\cdot\vec{E}_1\right|^2 + 2\left|\vec{E}_1\cdot\vec{E}^*_1\right|^2 + 2(\vec{E}_1\cdot{\vec{E}^*_1})(\vec{E}_3\cdot\vec{E}^*_3)\right.}{ \int d^3\vec{x} \hspace{0.02 in} \varepsilon \left|\vec{E}_1\right|^2}\right. \nonumber \\
 & & + \left.\frac{\left.2\left|\vec{E}_1\cdot\vec{E}_3\right|^2 + 2\left|\vec{E}_1\cdot\vec{E}^*_3\right|^2 + 3(\vec{E}^*_1\cdot{\vec{E}^*_1})(\vec{E}^*_1\cdot{\vec{E}_3}) \hspace{0.02 in} \right)}{ \int d^3\vec{x} \hspace{0.02 in} \varepsilon \left|\vec{E}_1\right|^2} \right]
\end{eqnarray}

\begin{eqnarray}
  \frac{\delta\omega_3}{\omega_3} &=& -\frac{1}{8}\left[ \frac{\int d^3\vec{x} \varepsilon \chithree \hspace{0.02 in} \left( \left|\vec{E}_3\cdot\vec{E}_3\right|^2 + 2\left|\vec{E}_3\cdot\vec{E}^*_3\right|^2 + 2(\vec{E}_1\cdot{\vec{E}^*_1})(\vec{E}_3\cdot\vec{E}^*_3)\right.}{ \int d^3\vec{x} \hspace{0.02 in} \varepsilon \left|\vec{E}_3\right|^2}\right. \nonumber \\
 & & + \left.\frac{\left.2\left|\vec{E}_1\cdot\vec{E}_3\right|^2 + 2\left|\vec{E}_1\cdot\vec{E}^*_3\right|^2 + (\vec{E}_1\cdot{\vec{E}_1})(\vec{E}_1\cdot{\vec{E}^*_3}) \hspace{0.02 in} \right)}{ \int d^3\vec{x} \hspace{0.02 in} \varepsilon \left|\vec{E}_3\right|^2} \right]
\end{eqnarray}
\end{widetext}
There is a subtlety in the application of perturbation theory to
decaying modes, such as those of a cavity coupled to output ports. In
this case, the modes are not truly eigenmodes, but are rather ``leaky
modes''~\cite{Snyder83}, and are not normalizable.  Perturbative
methods in this context are discussed in more detail
by~\cite{Suh04,Snyder83}, but for a tightly confined cavity mode it is
sufficient to simply ignore the small radiating field far away from
the cavity.  The field in the cavity is very nearly that of a true
eigenmode of an isolated cavity.

As stated above, we can arrive at the coupling coefficients by setting
$\omega_k \rightarrow \omega_k + \delta\omega_k$ in
Eq.~\ref{eq1}. However, the frequency perturbations
$\delta\omega_k$ are time-independent quantities, and we need to
connect them to the time-dependent $a_k$ amplitudes. Therefore, to
re-introduce the time dependence, one can use the slowly varying
envelope approximation: a slowly varying, time-dependent amplitude
$a_k(t)$ is introduced into the unperturbed fields $\vec{E}_k
\rightarrow \vec{E}_k a_k(t)$. The eigenmode must be normalized so
that $|a_k|^2$ is the energy, as assumed for the coupled-mode theory.
Thus, we divide each $\vec{E}_k$ by $\sqrt{\frac{1}{2} \int
\varepsilon |\vec{E}_k|^2}$.

First, we consider the $\chitwo$ medium.  Carrying out the above
substitutions in Eq.~\ref{eq1} and grouping terms proportional
$a_k$ yields Eqs.~\ref{chi2a}--\ref{chi2b} with $\alpha_{ij}$ and
$\beta_i$ given by:

\begin{align}
  \alpha_{ij} &= 0 \label{alpha2} \\ 
  \beta_1 &= \frac{1}{4} \frac{\int d^3\vec{x} \hspace{0.05 in}
    \sum_{ijk} \varepsilon \chitwo_{ijk} \hspace{0.02
    in}\left[\E{1}{i}^*\big(\E{2}{j}\E{1}{k}^*+\E{1}{j}^*\E{2}{k}\big)
    \right]}{\left[\int d^3\vec{x} \hspace{0.05 in} \varepsilon
    \left|\vec{E}_1\right|^2\right] \left[\int d^3\vec{x} \hspace{0.05 in} \varepsilon \left|\vec{E}_2\right|^2\right]^{1/2}} \\
  \beta_2 &= \frac{1}{4}\frac{\int d^3\vec{x}
    \hspace{0.05 in} \sum_{ijk} \varepsilon \chitwo_{ijk} \hspace{0.02 in}
    \E{2}{i}^*\E{1}{j}\E{1}{k}}{\left[\int d^3\vec{x} \hspace{0.05 in} \varepsilon
    \left|\vec{E}_1\right|^2\right] \left[\int d^3\vec{x} \hspace{0.05 in} \varepsilon \left|\vec{E}_2\right|^2\right]^{1/2}} 
\end{align}

A similar calculation yields the $\chithree$ coupled-mode equations
with coefficients given by:

\begin{align}
    \alpha_{ii} &= \frac{1}{8} \frac{\int d^3\vec{x} \hspace{0.05 in}
      \varepsilon \chithree \left|\vec{E}_i\cdot\vec{E}_i\right|^2 + \left|\vec{E}_i\cdot\vec{E}^*_i\right|^2}{\left[ \int d^3\vec{x}
	\hspace{0.05 in} \varepsilon
	\left|\vec{E}_i\right|^2\right]^2} \\ \alpha_{31} &=
	\frac{1}{4} \frac{\int d^3\vec{x} \hspace{0.05 in} \varepsilon \chithree
	\left|\vec{E}_1\right|^2\left|\vec{E}_3\right|^2 +
	\left|\vec{E}_1\cdot\vec{E}_3\right|^2 +
	\left|\vec{E}_1\cdot\vec{E}^*_3\right|^2}{\left[\int d^3\vec{x}
	\hspace{0.05 in} \varepsilon \left|\vec{E}_1\right|^2 \right]
	\left[\int d^3\vec{x} \hspace{0.05 in} \varepsilon
	\left|\vec{E}_3\right|^2 \right]} \\ \alpha_{13} &=
	\alpha_{31} \nonumber \\ \beta_1 &= \frac{3}{8} \frac{\int
	d^3\vec{x}
      \hspace{0.05 in} \varepsilon \chithree
      (\vec{E}^*_1\cdot\vec{E}^*_1)(\vec{E}^*_1\cdot\vec{E}_3)}{\left[\int
	d^3\vec{x} \hspace{0.05 in} \varepsilon
	\left|\vec{E}_1\right|^2\right]^{3/2} \left[\int d^3\vec{x}
	\hspace{0.05 in} \varepsilon
	\left|\vec{E}_3\right|^2\right]^{1/2}} \\
    \beta_3 &=
    \frac{1}{8} \frac{\int d^3\vec{x} \hspace{0.05 in} \varepsilon \chithree
      (\vec{E}_1\cdot\vec{E}_1)(\vec{E}_1\cdot\vec{E}^*_3)}{\left[\int
	d^3\vec{x} \hspace{0.05 in} \varepsilon
	\left|\vec{E}_1\right|^2\right]^{3/2} \left[\int d^3\vec{x}
	\hspace{0.05 in} \varepsilon
	\left|\vec{E}_3\right|^2\right]^{1/2}}
    \label{coeff3}
\end{align}

Note that Eqs.~\ref{alpha2}--\ref{coeff3} verify the conditions
$\omega_1\beta_1 = \omega_2\beta^*_2$ and $\omega_1\beta_1 =
\omega_3\beta^*_3$, previously derived from conservation of
energy---for $\chitwo$, this requires that one apply the symmetries of
the $\chitwo_{ijk}$ tensor, which is invariant under permutations of
$ijk$ for a frequency-independent
$\chitwo$~\cite{Boyd92}. Furthermore, we can relate the coefficients
$\alpha$ and $\beta$ to an effective modal volume $V$, similar to
\citeasnoun{Soljacic02:bistable}.  In particular, the strongest
possible nonlinear coupling will occur if the eigenfields are a
constant in the nonlinear material and zero elsewhere.  In this case,
any integral over the fields will simply yield the geometric volume
$V$ of the nonlinear material. Thus, for the $\chitwo$ effect we would
obtain $\beta_i \sim \chitwo / \sqrt{V\varepsilon}$; similarly, for
the $\chithree$ effect we would obtain $\alpha_{ij}, \beta_i \sim
\chithree / V \varepsilon$.  This proportionality to $1/\sqrt{V}$ and
$1/V$ carries over to more realistic field profiles (and in fact could
be used to \emph{define} a modal volume for these effects).

\section{Numerical validation}

To check the predictions of the $\chithree$ coupled-mode equations, we
performed an FDTD simulation of the one-dimensional waveguide-cavity
system shown in Fig~\ref{fig1}, whose analytical properties are
uniquely suited to third-harmonic generation. (The FDTD method,
including techniques to simulate nonlinear media, is described in
\citeasnoun{Taflove00}.)  This geometry consists of a semi-infinite
photonic-crystal structure made of alternating layers of dielectric
($\varepsilon_1=13$ and $\varepsilon_2 = 1$) with period $a$ and
thicknesses given by the quarter-wave condition ($d_1 = \sqrt
\varepsilon_2 / (\sqrt \varepsilon_1+ \sqrt \varepsilon_2)$ and $d_2 =
a-d_1$, respectively).  Such a quarter-wave stack possesses a periodic
sequence of photonic band gaps centered on frequencies $\omega_1 =
(\sqrt \varepsilon_1+ \sqrt \varepsilon_2) / 4\sqrt{\varepsilon_1
\varepsilon_2}$ (in units of $2\pi c/a$) for the lowest gap, and
higher-order gaps centered on odd multiples of $\omega_1$.  Moreover,
a defect formed by doubling the thickness of a $\varepsilon_1$ layer
creates cavity modes at exactly the middle of every one of these gaps.
Therefore, it automatically satisfies the frequency-matching condition
for third-harmonic generation.  In fact, it is too good: there will
also be ``ninth harmonic'' generation from $\omega_3$ to $\omega_9$.
This unwanted process is removed, however, by the discretization error
of the FDTD simulation, which introduces numerical dispersion that
shifts the higher-frequency modes.  To ensure the $\omega_3 =
3\omega_1$ condition in the face of this dispersion, we slightly
perturbed the structure (increasing the dielectric constant slightly
at the nodes of the third-harmonic eigenfield) to tune the
frequencies.  The simulated crystal was effectively semi-infinite,
with many more layers on the right than on the left of the cavity.  On
the left of the cavity, after two period of the crystal the material
is simply air ($\varepsilon=1$), terminated by a perfectly matched
layer (PML) absorbing boundary region.

We excite the cavity with an incident plane wave of frequency
$\omega_1$, and compute the resulting reflection spectrum.  The
reflected power at $\omega_3$, the third-harmonic generation, was then
compared with the prediction of the coupled-mode theory.  The
frequencies, decay rates, and $\alpha$ and $\beta$ coefficients in the
coupled-mode theory were computed from a linear FDTD simulation in
which the eigenmodes were excited by narrow-band pulses.  The freely
available FDTD code of~\cite{Farjadpour06} was employed.

\begin{figure}[htbp]
\centering\includegraphics[width=8cm]{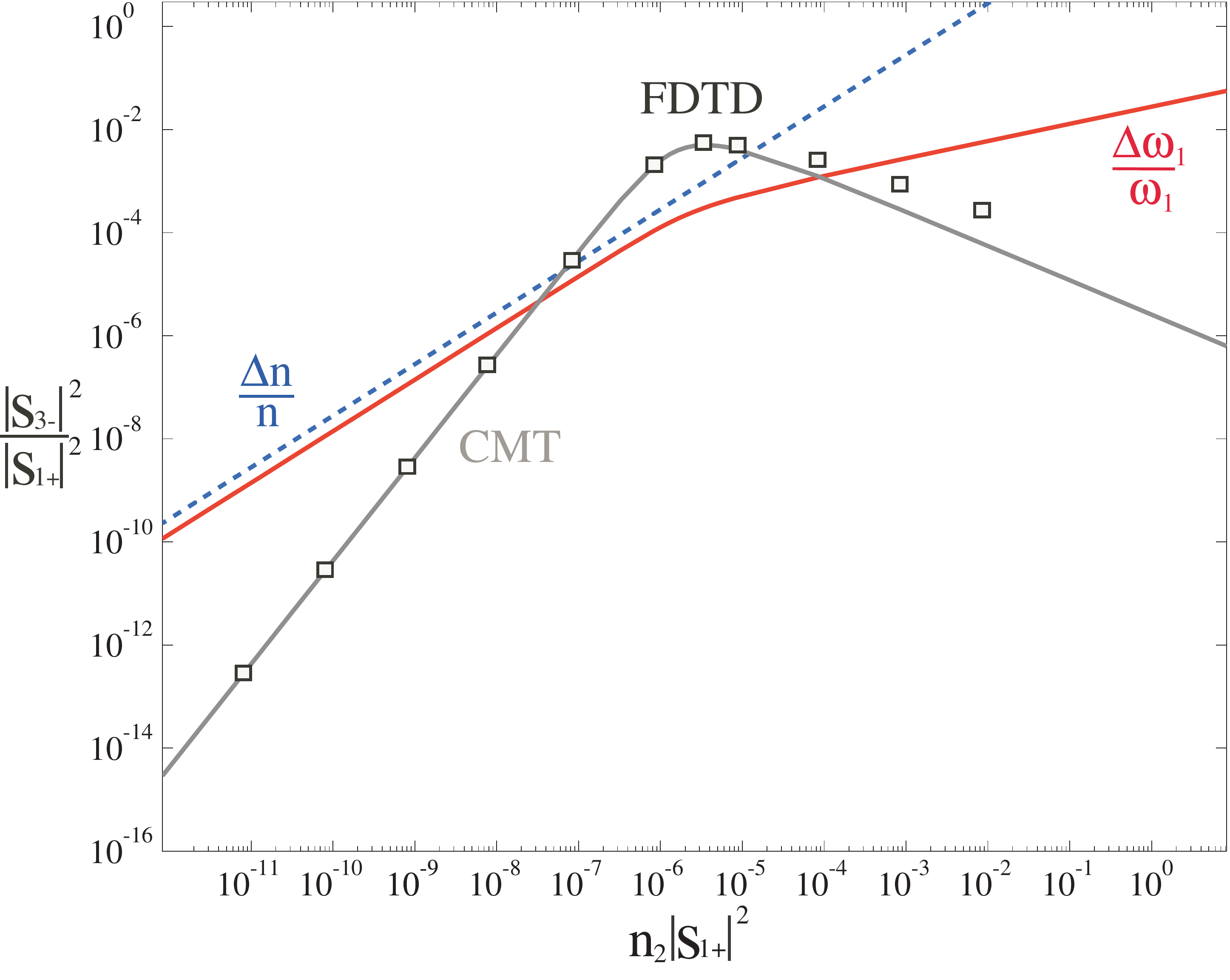}
\caption{Log-log plot of $|s_{3-}|^2/|s_{1+}|^2$ vs. $n_2 |s_{1+}|^2$
  for the coupled-mode theory (grey) and FDTD (black squares), where
  $n_2$ is being varied. Also shown are the corresponding $\Delta n/n$
  (dashed blue) and $\Delta \omega_1/\omega_1$ (solid red) curves.}
\label{fig2a}
\end{figure}

The results are shown in Fig.~\ref{fig2a}, in which the output power
at $\omega_1$ and $\omega_3 = 3\omega_1$ is denoted by $|s_{1-}|^2$
and $|s_{3-}|^2$, respectively, while the input power at $\omega_1$ is
denoted by $|s_{1+}|^2$.  In particular, we plot convenient
dimensionless quantities: the third-harmonic conversion efficiency
$\left|s_{3-}\right|^2 / \left|s_{1+}\right|^2$ as a function of the
dimensionless product $n_2 \left|s_{1+}\right|^2$ in terms of the
standard Kerr coefficient $n_2 = 3\chithree/4c\varepsilon$.  There is
clear agreement between the FDTD and CMT for small values of
$n_2\left|s_{1+}\right|^2$ (in which limit the conversion goes
quadratically with $n_2\left|s_{1+}\right|^2$).  However, as the input
power increases, they eventually begin to disagree, marking the point
where second-order corrections are required. This disagreement is not
a practical concern, however, because the onset of second-order
effects coincides with the limits of typical materials, which usually
break down for $\Delta n/n \equiv \chithree \text{max}|E|^2 /
2\varepsilon > 1\%$.  This is why we also plot the maximum index shift
$\Delta n/n$ in the same figure.

Also shown in Fig.~\ref{fig2a} is a plot of $\Delta\omega_1 /\omega_1
= \Re[\delta\omega_1/\omega_1]$. As expected, when $\Delta\omega_1$ is
of the order of $1/Q_1 \sim 10^{-3}$, the frequency shift begins to
destroy the frequency matching condition, substantially degrading the
third-harmonic conversion. (It might seem that $\Delta n /n$ and
$\Delta\omega_1 / \omega_1$ should be comparable, but this is not the
case because $\Delta n /n$ is the maximum index shift while
$\Delta\omega_1 / \omega_1$ is due to an average index shift.)

More specifically, the details of our simulation are as follows.  To
simulate a continuous wave (CW) source spectrum in FDTD, we employ a
narrow-bandwidth gaussian pulse incident from the air region, which
approximates a CW source in the limit of narrow bandwidth. This pulse
is carefully normalized so that the peak \emph{intensity} is unity, to
match the CMT.  The field in the air region is Fourier transformed and
subtracted from the incident field to yield the reflected flux.  Using
only two periods of quarter-wave stack on the left of the cavity we
obtained two cavity modes with real frequencies $\omega_1 = 0.31818$
(2$\pi$c/a), $\omega_2 = 0.95454$ (2$\pi$c/a) and quality factors $Q_1
= 1286$ and $Q_3 = 3726$, respectively. Given these field patterns, we
computed the $\alpha_{ij}$ and $\beta_i$ coefficients. We obtained the
following coupling coefficients, in units of $\chithree/a$:
$\alpha_{11} = 4.7531\times 10^{-4}, \alpha_{22} = 5.3306 \times
10^{-4}, \alpha_{12} = \alpha_{21} = 2.7847\times 10^{-4}, \beta_1 =
(4.55985 - 0.7244i)\times 10^{-5}$.

\section{Complete frequency conversion}

We now consider the conditions under which one may achieve
\emph{complete} frequency conversion: 100\% of the incident power
converted to output at the second or third harmonic frequency.  As we
shall see, this is easiest to achieve in the $\chitwo$ case, and
requires additional design criteria in the $\chithree$ case.

The key fact in a $\chitwo$ medium is that there are no
frequency-shifting terms ($\alpha = 0$), so the resonance condition
$\omega_2 = 2 \omega_1$ is not spoiled as one increases the power.
The only requirement that we must impose is that external losses such
as absorption are negligible ($\tauk{e}{k} \gg \tauk{s}{k}$).  In this
case, 100\% conversion corresponds to setting $s_{1-} = 0$ in the
steady-state.  Using this fact, Eqs.~\ref{chi2a}-\ref{chi2b} for an
input source $s_{+}(t) = s_{1+}\exp(i\omega_1t)$ yields the following
condition on the input power for 100\% conversion:
\begin{equation}
 \left|s_{1+}\right|^2
= \frac{2}{\omega^2_1\left|\beta_1\right|^2\tauk{s}{2}\tauk{s}{1}^2} 
= \frac{\omega_1}{2\left|\beta_1\right|^2 Q_2 Q_1^2}
\label{effconv2}
\end{equation}
(A similar dependence of efficiency on $Q_1^2 Q_2$ was previously
observed~\cite{Berger96,Liscidini06}, although a critical power was
not identified.) Thus, we can always choose an input power to obtain
100\% conversion.  If $Q_1 \sim Q_2$, then this critical power scales
as $V/Q^3$ where $V$ is the modal volume (recall that $\beta \sim
1/\sqrt{V}$).

This is limited, however, by our first-order approximation in
perturbation theory: if the input power becomes so large that
second-order effects (or material breakdown) become significant, then
this prediction of 100\% conversion is no longer valid.  The key
condition is that the fractional change in the refractive index be
small: $\Delta n / n \ll 1$.  This can always be satisfied, in
principle: if one chooses $Q_1$ and/or $Q_2$ to be sufficiently large,
then the critical power can be made arbitrarily small in
principle. Not only does the critical power decrease with $Q^3$, but
the field intensity in the cavity ($|a_i|^2$) decreases as $V/Q_1
Q_2$, and thus one can avoid large $\Delta n / n$ as well as lowering
the power.  (Note that the field intensity goes as $1/Q^2$ while the
power goes as $1/Q^3$ simply because the energy and power are related
by a time scale of $Q$.)

\begin{figure}[hbt]
\centering\includegraphics[width=7cm]{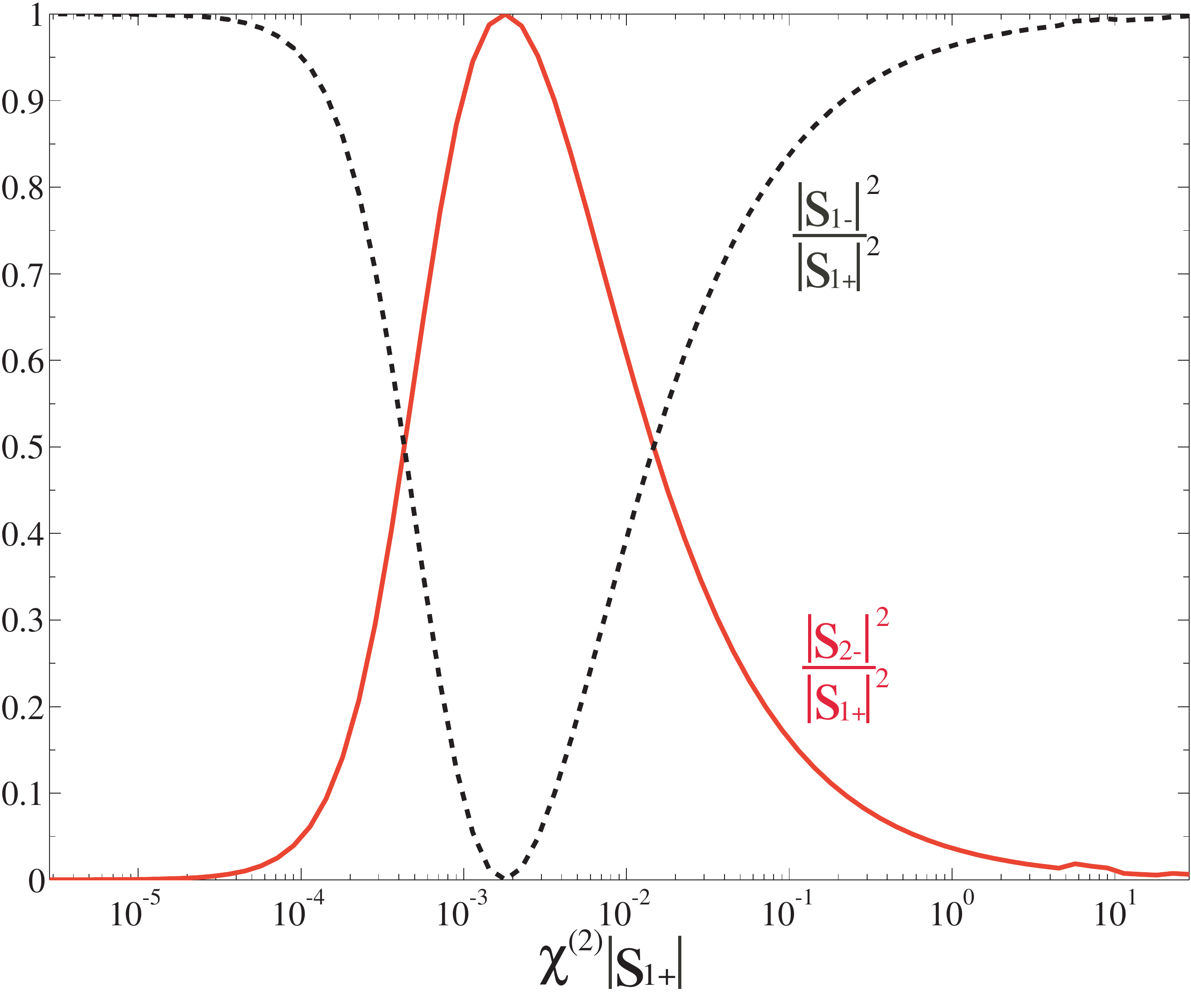}
\caption{Plot of first and second harmonic efficiency,
  $|s_{1-}|^2/|s_{1+}|^2$ (black) and $|s_{2-}|^2/|s_{1+}|^2$ (red),
  vs. $\chitwo |s_{1+}|$. $100\%$ power transfer from $\omega_1$ to
  $\omega_2 = 2\omega_1$ is achieved at $\chitwo |s_{1+}| =
  1.8\times10^{-3}$.}
\label{fig3}
\end{figure}

To illustrate second-harmonic conversion for a $\chitwo$ medium, we
plot the solution to the coupled-mode equations as a function of input
power in Fig.~\ref{fig3}.  The 100\% conversion at the predicted
critical power is clearly visible.  For this calculation, we chose
modal parameters similar to the ones from the FDTD computation before:
$\omega_1 = 0.3$, $\omega_2 = 0.6$, $Q_1 = 10^4$, $Q_2 = 2\times10^4$,
with dimensionless $\beta_1 = (4.55985 - 0.7244)\times 10^{-5}$.

A $\chithree$ medium, on the other hand, does suffer from nonlinear
frequency shifts.  For example, Fig.~\ref{fig2a}, which is by no means
the optimal geometry, exhibits a maximal efficiency of
$|s_{3-}|^2/|s_{1+}|^2 \approx 4 \times 10^{-3}$, almost three orders
of magnitude away from complete frequency conversion.  On the other
hand, we can again achieve 100\% conversion if we can force
$\alpha_{ij} = 0$, which can be done in two ways.  First, one could
employ \emph{two} $\chithree$ materials with opposite-sign $\chithree$
values (e.g., as in \citeasnoun{Smith97}). For example, if the
$\chithree$ is an \emph{odd} function around the cavity center, then
the integrals for $\alpha_{ij}$ will vanish while the $\beta$
integrals will not.  (In practice, $\alpha \ll \beta$ should suffice.)
Second, one could pre-compensate for the nonlinear frequency shifts:
design the cavity so that the shifted frequencies, at the critical
power below, satisfy the resonant condition $\omega_3 + \Delta\omega_3
= 3 (\omega_1 + \Delta\omega_1)$.  Equivalently, design the device for
$\alpha_{ij}=0$ and then adjust the linear cavity frequencies {\it a
posteriori} to compensate for the frequency shift at the critical
power.  (This is closely analogous to the cavity detuning used for
optical bistability~\cite{Soljacic02:bistable}, in which one operates
off-resonance in the linear regime so that resonance occurs from the
nonlinear shift.)

If $\alpha_{ij}$ is thereby forced to be zero, and we can also
neglect external losses (absorption, etc.) as above, then 100\%
third-harmonic conversion ($s_{1-}=0$) is obtained when:
\begin{equation}
 \left|s_{1+}\right|^2
= \left[
  \frac{4}{\omega^2_1\left|\beta_1\right|^2\tauk{s}{1}^3\tauk{s}{3}}
  \right]^{1/2}
= \left[
  \frac{\omega_1 \omega_3}{4\left|\beta_1\right|^2 Q_1^3 Q_3}
  \right]^{1/2}
\label{effconv3}
\end{equation}
If $Q_1 \sim Q_3$, then this critical power scales as $V/Q^2$ where
$V$ is the modal volume (recall that $\beta \sim 1/V$).  This is
precisely the scaling that was predicted for the power to obtain
nonlinear bistability in a single-mode cavity~\cite{Yanik04}.
Similarly, one finds that the energy density in the cavity ($|a_i|^2$)
decreases proportional to $V/\sqrt{Q_1 Q_3}$.

\begin{figure}[hbt]
\centering\includegraphics[width=7cm]{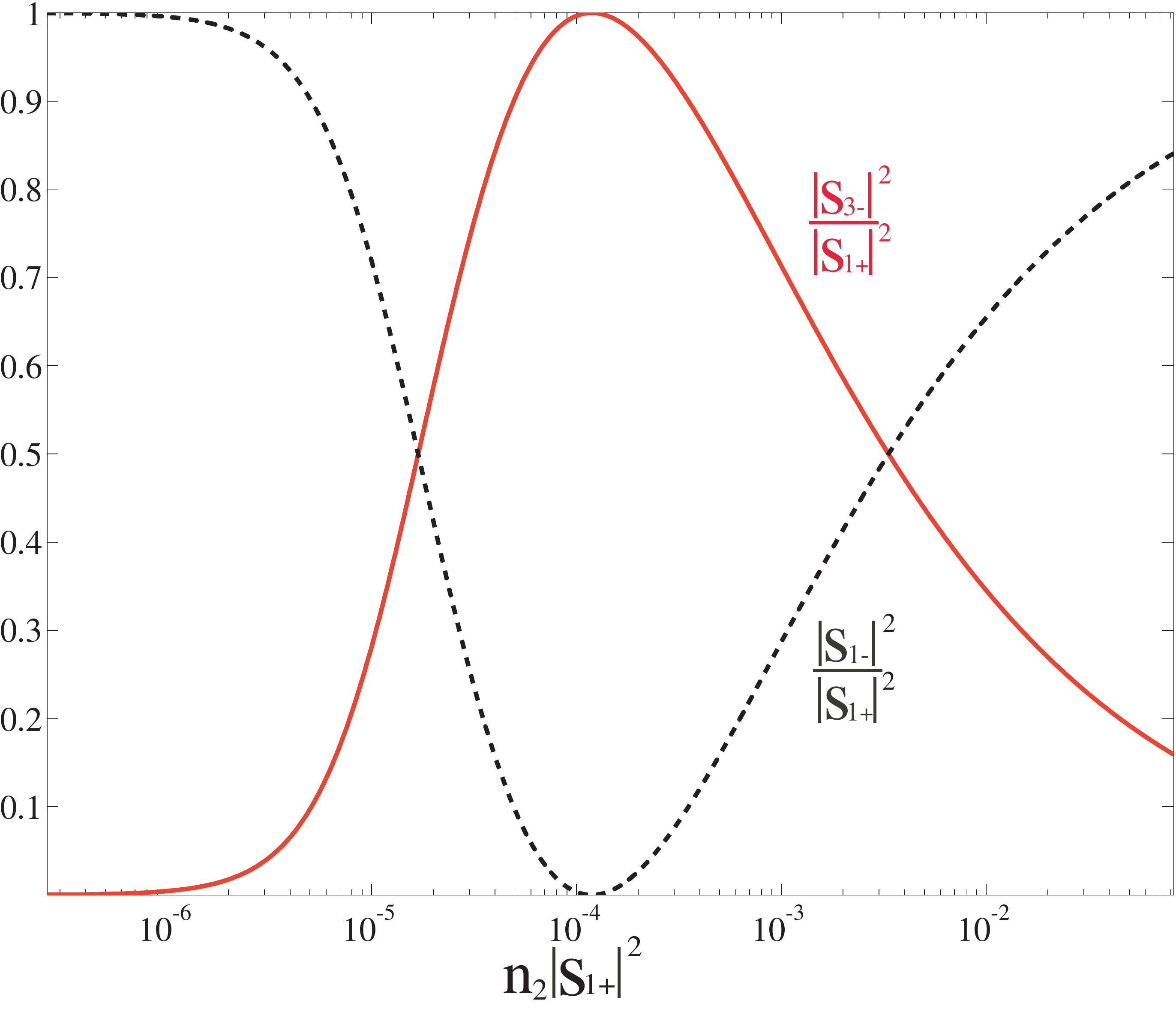}
\caption{Plot of first and third harmonic efficiency,
  $|s_{1-}|^2/|s_{1+}|^2$ (black) and $|s_{3-}|^2/|s_{1+}|^2$ (red),
  vs. $n_2 |s_{1+}|^2$. $100\%$ power transfer from $\omega_1$ to
  $\omega_3 = 3\omega_1$ is achieved at $n_2 |s_{1+}|^2 = 2.8 \times
  10^{-4}$.}
\label{fig4}
\end{figure}

We demonstrate the third-harmonic conversion for $\alpha_{ij}=0$ by
plotting the solution to the coupled-mode equations as a function of
input power in Fig.~\ref{fig4}.  Again, 100\% conversion is only
obtained at a single critical power.  Here, we used the same
parameters as in the FDTD calculation, but with $\alpha = 0$.  In this
case, comparing with Fig.~\ref{fig2a}, we observe that complete
frequency conversion occurs at a power corresponding to $\Delta n/n
\approx 10^{-2}$.  This is close to the maximum power before
coupled-mode/perturbation theory becomes invalid (either because of
second-order effects or material breakdown), but we could easily
decrease the critical power by increasing $Q$.

For both the $\chitwo$ and the $\chithree$ effects, in
Figs.~\ref{fig3}--\ref{fig4}, we see that the harmonic conversion
efficiency goes to zero if the input power (or $\chi$) is either too
small or too large.  It is not surprising that frequency conversion
decreases for low powers, but the decrease in efficiency for high
powers is less intuitive.  It corresponds to a well-known phenomenon
in coupled-mode systems: in order to get 100\% transmission from an
input port to an output port, the coupling rates to the two ports must
be matched in order to cancel the back-reflected
wave~\cite{Haus84:coupled, Fan01}. In the present case, the coupling
rate to the input port is $\sim 1/Q_1$, and the coupling rate to the
output ``port'' (the harmonic frequency) is determined by the strength
of the nonlinear coupling.  If the nonlinear coupling is either too
small or too large, then the rates are not matched and the light is
reflected instead of converted.  (On the other hand, we find that for
large input powers, while the conversion \emph{efficiency} as a
fraction of input power goes to zero, the \emph{absolute} converted
power ($|s_{2-}|^2$ or $|s_{3-}|^2$) goes to a constant.)

Finally, let us consider one other potential problem.  Any physical
$\chitwo$ medium will generally also have $\chithree \neq 0$, so if
the power is large enough this could conceivably cause a frequency
shift that would spoil the second-harmonic resonance even in the
$\chitwo$ device.  Here, we perform a simple scaling analysis to
determine when this will occur. (Although the frequency shifting could
potentially be compensated for as described above, one prefers that it
be negligible to begin with.)  In order to preserve the resonance
condition, any fractional frequency shift $\Delta\omega/\omega$ must
be much smaller than the bandwidth $1/Q$, or equivalently we must have
$Q\Delta\omega/\omega \ll 1$. From above, $\Delta\omega \sim \omega
\alpha |a|^2$, and $|a|^2 \sim |s_{1+}|^2 Q/\omega$.  Suppose that we
are operating at the critical input power $P^{(2)}$ for
second-harmonic conversion, from \eqref{effconv2}.  It then follows
that we desire $Q\Delta\omega/\omega \sim Q^2 \alpha/\omega P^{(2)}
\ll 1$.  It is convenient to re-express this relationship in terms of
$P^{(3)} \sim \omega / \beta Q^2$, the third-harmonic critical power
from \eqref{effconv2}, by assuming $\alpha \sim \beta$ as discussed in
the previous section.  We therefore find that $\chithree$ self-phase
modulation can be ignored for $\chitwo$ second-harmonic generation as
long as $P^{(2)} / P^{(3)} \ll 1$.  As discussed in the concluding
remarks, this is indeed the case for common materials such as gallium
arsenide, where $P^{(2)} / P^{(3)} \approx 1/30$ for $Q \sim 1000$ and
for typical values of the cavity lifetime and volume.  Moreover, since
$P^{(2)} / P^{(3)} \sim 1/Q$, one can make the ratio arbitrarily
smaller if necessary (at the expense of bandwidth) by increasing $Q$.

\section{The Effect of Losses}
\label{sec:loss}

In practice, a real device will have some additional losses, such as
linear or nonlinear absorption and radiative scattering.  Such losses
will lower the peak conversion efficiency below 100\%.  As we show in
this section, their quantitative effect depends on the ratio of the
loss rate to the total loss rate $1/Q$.  We also solve for the
critical input power to achieve maximal conversion efficiency in the
presence of losses.

For a $\chitwo$ medium with a linear loss rate $1/\tauk{e}{k}$, we
solve Eqs~\ref{chi2a}--\ref{chi2b} for $|s_{2-}|^2$ and enforce the
condition for maximal conversion efficiency:
$\frac{d}{d|s_{1+}|^2}(|s_{2-}|^2/|s_{1+}|^2) = 0$.  We thus obtain
the following optimal input power and conversion efficiency:
\begin{align}
 \left|s_{1+}\right|^2 &=
 \frac{2\tauk{s}{1}}{\omega^2_1\left|\beta_1\right|^2\tau^3_1\tau_2}
 \label{abs1} \\ \frac{\left|s_{2-}\right|^2}{\left|s_{1+}\right|^2}
 &= \frac{\tau_1\tau_2}{\tauk{s}{1}\tauk{s}{2}} \label{abs2}
\end{align}
It immediately follows that for zero external losses, i.e. $\tau_k =
\tauk{s}{k}$, Eq.~\ref{abs2} gives 100\% conversion and Eq.~\ref{abs1}
reduces to Eq.~\ref{effconv2}. For small external losses $\tauk{s}{k}
\ll \tauk{e}{k}$, the optimal efficiency is reduced by the ratio of
the loss rates, to first order:
\begin{equation}
  \frac{\left|s_{2-}\right|^2}{\left|s_{1+}\right|^2} \approx 1- \left(
  \frac{\tauk{s}{2}}{\tauk{e}{2}} +
  \frac{\tauk{s}{1}}{\tauk{e}{1}} \right).
  \label{effabs}
\end{equation}
(A similar transmission reduction occurs in coupled-mode theory when
any sort of loss is introduced into a resonant coupling
process~\cite{Haus84}.)

The same analysis for $\chithree$ yields the following critical
input power and optimal efficiency:

\begin{align}
 \left|s_{1+}\right|^2 &=
 \left[\frac{4\tauk{s}{1}^2}{\omega^2_1\left|\beta_1\right|^2\tau^5_1\tau_3}\right]^{1/2}
 \label{abs1chi3} \\ \frac{\left|s_{3-}\right|^2}{\left|s_{1+}\right|^2}
 &= \frac{\tau_1\tau_3}{\tauk{s}{1}\tauk{s}{3}}
 \label{abs2chi3}
\end{align}
where by comparison with Eq.~\ref{abs2}, a first-order expansion for
low-loss yields an expression of the same form as Eq.~\ref{effabs}:
the efficiency is reduced by the ratio of the loss rates, with
$\tau_2$ replaced by $\tau_3$.

A $\chithree$ medium may also have a nonlinear ``two-photon''
absorption, corresponding to a complex-valued $\chithree$, which gives
an absorption coefficient proportional to the field intensity.  This
enters the coupled-mode equations as a small imaginary part added to
$\alpha$, even if we have set the real part of $\alpha$ to zero.  (The
corresponding effect on $\beta$ is just a phase shift.)  That yields a
nonlinear (NL) $\tauk{e}{k}$ of the following form, to lowest order in the loss:
\begin{align}
\frac{1}{\tauk{e}{1}^\textrm{NL}} &\approx \omega_1 \Im \left[ \alpha_{11} \frac{\tauk{s}{1}}{2} |s_{1+}|^2
                   + \alpha_{13} \frac{\tauk{s}{3}^2 \tauk{s}{1}^3}{8} \omega_3^2 |\beta_3|^2 |s_{1+}|^6 \right]
\\
\frac{1}{\tauk{e}{3}^\textrm{NL}} &\approx  \omega_3 \Im \left[ \alpha_{31} \frac{\tauk{s}{1}}{2} |s_{1+}|^2
                   + \alpha_{33} \frac{\tauk{s}{3}^2 \tauk{s}{1}^3}{8} \omega_3^2 |\beta_3|^2 |s_{1+}|^6 \right] .
\end{align}
where we have simply substituted the values for the critical fields
$a_1= \sqrt{2/\tau_1} s_{1+}$ and $a_3$ given by \eqref{chi3b}, and
grouped terms that correspond to imaginary frequency shifts. These
loss rates can then be substituted in the expression for the losses
above, i.e. \eqref{abs2chi3}, in which case one obtains the following
optimal efficiency of third-harmonic generation, to lowest-order in
the loss, not including linear losses:
\begin{equation}
  \frac{\left|s_{3-}\right|^2}{\left|s_{1+}\right|^2} \approx 1 - \frac{\tauk{s}{3}}{\left|\beta_1\right|} \sqrt{\frac{\tauk{s}{3}}{\tauk{s}{1}}} \Im \left[\frac{\alpha_{11}+3\alpha_{13}}{\tauk{s}{3}} +
  \frac{\alpha_{13}+3\alpha_{33}}{\tauk{s}{1}}\right]
\label{effabsNL}
\end{equation}
(The linear and nonlinear losses can be combined by simply multiplying
Eq.~\ref{abs2chi3} and Eq.~\ref{effabsNL}.)  Thus, the nonlinear loss
is proportional to the ratio $\Im \alpha/ |\beta|$, which is
proportional to $\Im \chithree / |\chithree|$.

\section{Conclusion}

We have presented a rigorous coupled-mode theory for second- and
third-harmonic generation in doubly resonant nonlinear cavities,
accurate to first order in the nonlinear susceptibility and validated
against a direct FDTD simulation.  Our theory, which generalizes
previous work on this subject, predicts several interesting
consequences.  First, it is possible to design the cavity to yield
100\% frequency conversion in a passive (gain-free) device, even when
nonlinear down-conversion processes are included, limited only by
fabrication imperfections and losses.  Second, this 100\% conversion
requires a certain critical input power---powers either too large or
too small lead to lower efficiency.  Third, we describe how to
compensate for the self-phase modulation in a $\chithree$ cavity.  The
motivation for this work was the hope that a doubly resonant cavity
would lead to 100\% conversion at very low input powers, and so we
conclude our paper by estimating the critical power for reasonable
material and geometry assumptions.

A typical nonlinear material is gallium arsenide (GaAs), with $\chitwo
\approx 145$ $\text{pm}/\text{V}$ and $n_2 = 1.5 \times 10^{-13}$
$\text{cm}^2/\text{W}$ at $1.5 \mu\textrm{m}$.  (Al doping is usually
employed to decrease nonlinear losses near resonance~\cite{Villeneuve93}.)
Although this has both $\chitwo$ and $\chithree$ effects, we can
selectively enhance one or the other by choosing the cavity to have
resonances at either the second or third harmonic.  Many well confined
optical cavity geometries are available at these wavelengths and have
been used for nonlinear devices, such as ring resonators~\cite{Xu05}
or photonic-crystal slabs~\cite{Notomi05}.  We will assume
conservative parameters for the cavity: a lifetime $Q_1 = 1000$, $Q_2
= 2000$, $Q_3 = 3000$, and a modal volume of 10 cubic half-wavelengths
($V \approx 10(\lambda/2n)^3$) with roughly constant field amplitude
in the nonlinear material (worse than a realistic case of strongly
peaked fields).  In this case, the critical input power, from
Eqs.~\ref{effconv2}--\ref{effconv3}, becomes approximately 70~$\mu$W
for second-harmonic generation and 2~mW for third-harmonic generation
(with a moderate peak index shift $\Delta n/n \approx 10^{-3}$,
justifying our first-order approximation)

Future work will involve designing specific doubly resonant cavity
geometries and more precise power predictions.  Using our expressions
for $\alpha$ and $\beta$, optimized cavities for harmonic generation
can be designed using standard methods to compute the linear
eigenmodes.  In practice, experimentally achieving cavity modes with
``exactly'' harmonic frequencies, matched to within the fractional
bandwidth $1/Q$, is a challenge and may require some external tuning
mechanism.  For example, one could use the nonlinearity itself for
tuning, via external illumination of the cavity with an intense
``tuning'' beam at some other frequency.  Also, although we can
directly integrate the coupled-mode equations in time, we intend to
supplement this with a linearized stability analysis at the critical
power.  This is particularly important for the $\chithree$ case, where
pre-correcting the frequency to compensate the nonlinear frequency
shift (self-phase modulation) may require some care to ensure a stable
solution.

\section*{Acknowledgements}

We would like to thank Zheng Wang and Karl Koch for useful
discussions, as well as the anonymous referees for many helpful
suggestions. This work was supported in part by the Materials Research
Science and Engineering Center program of the National Science
Foundation under award DMR-9400334, by a Department of Energy (DOE)
Computational Science Fellowship under grant DE--FG02-97ER25308, and
also by the Paul E. Gray Undergraduate Research Opportunities Program
Fund at MIT.

\end{document}